\documentclass[12pt]{article}
\input epsf.sty
\usepackage{mathrsfs}
\usepackage{amsmath}
\usepackage{mathrsfs}
\usepackage{amssymb}
\usepackage{color}
\usepackage{psfrag}
\usepackage{graphicx}
\usepackage{subfigure}
\usepackage{rotating}

\newcommand{\bea}{\begin{eqnarray}}
\newcommand{\eea}{\end{eqnarray}}
\newcommand{\be}{\begin{equation}}
\newcommand{\ee}{\end{equation}}
\newcommand{\vs}[1]{\vspace{#1 mm}}

\newcommand{\dsl}{\pa \kern-0.5em /}

\newcommand{\half}{\frac{1}{2}}
\newcommand{\pa}{\partial}

\newcommand{\nn}{\nonumber\\}

\begin{document}
\topmargin 0pt
\oddsidemargin 0mm

\begin{flushright}

%USTC-ICTS-10-12\\

%hep-th/yymmnnn\\

\end{flushright}

\vspace{2mm}

\begin{center}
{\Large \bf Lifshitz metric with hyperscaling violation\\ 
from NS5-D$p$ states in string theory}

\vs{10}

{Parijat Dey\footnote{E-mail: parijat.dey@saha.ac.in} and 
Shibaji Roy\footnote{E-mail: shibaji.roy@saha.ac.in}}

 \vspace{4mm}

{\em

 Saha Institute of Nuclear Physics,
 1/AF Bidhannagar, Calcutta-700 064, India\\}

\end{center}

\vs{10}

\begin{abstract}
In previous papers \cite{Dey:2012tg, Dey:2012rs} we have shown 
how Lifshitz-like space-time (space-time
having a Lifshitz scaling along with a hyperscaling violation) can arise by
taking near horizon limits of certain intersecting solutions 
(F-string with D$p$-branes and also with two D-branes) of string theory. 
In this paper we construct intersecting bound state solutions in the form of 
NS5-D$p$-branes (with $1\leq p \leq 6$) of type II string theories. These are
1/4 BPS and threshold bound states unlike the known NS5-D$p$ bound states
which are 1/2 BPS and non-threshold. We show that the near horizon limits of
these solutions also lead to Lifshitz-like space-time with the dynamical
scaling exponent $z=0$ and the hyperscaling violation exponent $\theta=9-p$.
The spatial dimension of the boundary theory is $d=7-p$. The dilatons in these
theories are not constant in general (except for $p=5$) and therefore produce
RG flows. So, we also consider the strong coupling phases of these theories
and find that these phases also have similar Lifshitz-like structures,
except for $p=2$, where it has an AdS$_3$ structure.    
\end{abstract}
\newpage
\section{Introduction}

It is well-known that string theory admits various kinds of spatially extended
solutions, generically called branes, which preserve certain fraction of
space-time supersymmetries. These solutions may be constituted of a single
type of brane or a composite of more than one type of branes. They
have been proved quite useful as string theory in the near horizon 
geometry (a kind of low energy limit) of a stack of coincident branes can be 
seen to be holographically dual to a theory without gravity living 
on the boundary. This is a strong-weak duality and so one
can gain insight about the strongly coupled field theory by studying the
weakly coupled string or supergravity theory \cite{Maldacena:1997re,
Gubser:1998bc, Witten:1998qj, Aharony:1999ti}. For D3-brane this is the AdS/CFT
correspondence of Maladacena \cite{Maldacena:1997re} where the boundary 
theory is four
dimensional, ${\cal N}=4$, SU($N$) super Yang-Mills theory (which is 
conformal). Similar correspondence is also believed 
to hold for other types of branes which goes by the name gauge/gravity 
duality \cite{Aharony:1999ti}. So, for example, for D$p$-branes, 
the boundary theory is
$(p+1)$-dimensional super Yang-Mills theory (which is non-conformal) with 16 
supercharges \cite{Itzhaki:1998dd}.       

Similar correspondence holds good even for the composite brane states. 
Composite brane solutions also known as intersecting branes could be of 
different types. So, for example, there are 1/2 BPS non-threshold types, such
as, D$(p-2)$-D$p$ \cite{Russo:1996if, Breckenridge:1996tt,
Costa:1996zd} (for $1\leq p \leq 6$), F-D$p$ 
\cite{Lu:1999uca} (for $1\leq p \leq 7$) and 
NS5-D$p$ \cite{Lu:1998vh, Alishahiha:2000pu, Mitra:2000wr} 
(for $0\leq p \leq 5$) among others. 
In the near horizon limit (or decoupling limit) they give rise to some
non-gravitational, non-local theories on the boundary. For D$(p-2)$-D$p$, we
get noncommutative Yang-Mills (NCYM) theory in $(p+1)$-dimensions
\cite{Maldacena:1999mh, Hashimoto:1999ut}, for 
F-D$p$ we get $(p+1)$-dimensional noncommutative open string (NCOS) theory 
\cite{Seiberg:2000ms, Gopakumar:2000na} and for NS5-D$p$ we get six 
dimensional open D$p$-brane (OD$p$) theory \cite{Gopakumar:2000ep}. 
So, one can gain insight about these
non-local theories, in the strongly coupled regime, by studying the gravity
duals. The other type of composite brane solutions are the 1/4 BPS and 
threshold intersecting brane solutions \cite{Tseytlin:1997cs,
Gauntlett:1996pb, Behrndt:1996pm}. These are D$(p-4)$-D$p$ (for $3\leq p
\leq 6$) and we have also found some new solutions F-D$p$ (for $0\leq p \leq
5$) \cite{Dey:2012tg}, D$(p-2)$-D$p$ (for $3\leq p\leq 5$) and 
D$p$-D$p$$^{\prime}$ (for $2\leq p \leq 4$) \cite{Dey:2012rs}. All these
solutions (except some), as we have seen, give rise to Lifshitz-like space-time
in the near horizon limit\footnote{The concept of hyperscaling violation was
introduced in the context of random field Ising system in 
\cite{Fisher:1986zz}. In
gauge/gravity duality this was identified while describing certain metallic
states with hidden fermi surface in \cite{Ogawa:2011bz, Huijse:2011ef}. 
More general gravity solutions
having Lifshitz scaling alongwith hyperscaling violation have been found
and studied 
in \cite{Charmousis:2010zz, Gouteraux:2011ce, Shaghoulian:2011aa, 
Dong:2012se, Narayan:2012hk, Kim:2012nb,
Singh:2012un, Dey:2012tg, Dey:2012rs, Perlmutter:2012he, Cadoni:2012uf, 
Ammon:2012je, Bhattacharya:2012zu, Dey:2012hf, Kundu:2012jn, 
Alishahiha:2012he}.}. It is known 
that Lifshitz scaling 
symmetry, which is a non-relativistic symmetry, arise as a possible symmetry 
in some condensed matter systems at quantum critical point \cite{Sachdev,
Sachdev:2012dq}. So, it may be
the case that the space-time we obtain from these intersecting bound state
solutions in the near horizon limit are the gravity duals of such condensed
matter systems near quantum critical point and since the latter systems are
strongly coupled, we can learn about the phase structures by studying the
gravity solutions in the spirit of AdS/CFT correspondence.

In this paper we construct another kind of 1/4 BPS threshold intersecting
brane solutions of type II string theories in the form of NS5-D$p$ (for
$1 \leq p \leq 6$) solutions.
These solutions in the near horizon limit give rise to Lifshitz-like
space-time\footnote{Here in order not to confuse the readers we would
like to clarify that Lifshitz-like space-times actually arise after the
compactifications of the near horizon NS5-D$p$ metrics on S$^2$ $\times$
T$^{p-1}$ as described in the next section.}  and may describe the gravity 
dual of some condensed matter systems
near quantum critical point\footnote{Note that the number of transverse 
directions of the
NS5-D$p$-brane solutions we construct here is three for all $p$. So,
in analogy with D6-brane (where also the number of transverse directions is 
three), one might think that the gravity 
does not decouple \cite{Itzhaki:1998dd} in the near horizon limit of our 
solutions. Therefore, the near horizon limit of
NS5-D$p$-brane solutions may not describe the gravity dual of some condensed 
matter systems. However, this is not true and gravity does decouple for our
solutions in the near horizon limit. The simple reason is that
we have shown (later) that Lifshitz metric with hyperscaling
violation arise from the near horizon limit of our NS5-D$p$ brane solutions
after compactification on S$^2$ $\times$ T$^{p-1}$.
This geometry has been recognised \cite{Ogawa:2011bz, Huijse:2011ef, Dong:2012se} 
to be the gravity dual of some condensed
matter systems near the quantum critical point. So, as soon as we get this 
geometry (note that it is not obvious from which string theory solution
we can get this geometry, but obviously not from the solution where gravity
is known not to decouple), it automatically implies that gravity gets 
decoupled at least in lower dimensions. But since decoupling can not
occur just by dimesional reduction or compactification, the gravity does
decouple even in the ten dimensional near horizon NS5-D$p$ solutions.} 
(unlike the standard NS5-D$p$ which in the near 
horizon limit gives rise to OD$p$ theories \cite{Gopakumar:2000ep, 
Alishahiha:2000pu, Mitra:2000wr} , mentioned earlier). The
construction proceeds from the F-D5 solution we have obtained in
\cite{Dey:2012tg}. Then taking an S-duality on that solution gives 
NS5-D1 solution. Further
T-dualities along the spatial directions of NS5-brane yield all the NS5-D$p$
solutions which are 1/4 BPS and threshold solutions. Then taking the near 
horizon
limit and going to a suitable coordinate we get the Lifshitz space-time with
hyperscaling violation where the dynamical critical exponent we find is 
$z=0$ and the hyperscaling violation exponent is $\theta = 9-p$. The spatial
dimension of the boundary theory is given as $d=7-p$. These values of $(z,\,
\theta,\,d)$ are shown to satisfy the null energy condition (NEC)
\cite{Dong:2012se} so that they
may give sensible dual theories. We have seen that the dilatons for these 
solutions are in general non-constant except for $p=5$ and therefore produce RG
flows. We also construct the strongly coupled phases of these solutions
either by going to the
S-dual frame (for type IIB solutions) or by uplifting the solutions to
M-theory (for type IIA solutions). These strongly coupled phases also have
similar Lifshitz-like structures except for $p=2$, where the M-theory lift
of the NS5-D2 solution has AdS$_3$ structure in the near horizon limit.

This paper is organized as follows. In the next section, we give the
construction of NS5-D$p$ solutions of type II string theories and show that
their near horizon limit gives rise to Lifshitz-like space-time. In section 3
we discuss the RG flows and give the strongly coupled phases of the various
solutions obtained in the previous section. Then we conclude in section 4.

\section{NS5-D$p$ and Lifshitz-like space-time}

In this section we first construct the 1/4 BPS threshold NS5-D$p$ bound state
solution of type II string theories and then by taking the near horizon limit
we will show how they lead to Lifshitz-like space-time in a suitable
coordinate. For that we will start from the F-D5 solution obtained in eq.(2.6)
of \cite{Dey:2012tg}. For $p=5$, the F-D$p$ solution given there takes the
form, 
\bea\label{fd5}
ds^2 &=& H_2^{\half} \left[-\frac{dt^2}{H_1H_2}
+ \frac{\sum_{i=1}^5 (dx^i)^2}{H_2} + \frac{(dx^{6})^2}{H_1}+ dr^2 +
r^2 d\Omega_{2}^2\right]\nn
e^{2\phi} &=& \frac{1}{H_1H_2}\nn
B_{[2]} &=& \left(1-H_1^{-1}\right)dt \wedge dx^{6}, \qquad A_{[6]}\,\,
=\,\, \left(1-H_2^{-1}\right)dt\wedge dx^1 \wedge \cdots \wedge dx^5
\eea
Here the two harmonic functions are given as, $H_{1,2}=1+Q_{1,2}/r$, with
$Q_{1,2}$ denoting the charges of the F-string and D5-brane respectively.
It is clear from \eqref{fd5} that F-string is lying along $x^6$ and D5-brane
is lying along $x^1,\ldots, x^5$. Also $B_{[2]}$ is the NSNS form field which
couples to F-string and $A_{[6]}$ is the RR 6-form gauge field which couples 
to D5-brane. Taking an S-duality on this solution we get NS5-D1 solution and 
has the form,
\bea\label{ns5d1}
ds^2 &=& H_1^{\half}H_2 \left[-\frac{dt^2}{H_1H_2}
+ \frac{\sum_{i=1}^5 (dx^i)^2}{H_2} + \frac{(dx^{6})^2}{H_1}+ dr^2 +
r^2 d\Omega_{2}^2\right]\nn
e^{2\phi} &=& H_1H_2\nn
A_{[2]} &=& \left(1-H_1^{-1}\right)dt \wedge dx^{6}, \qquad H_{[3]}\,\,
=\,\, -Q_2 {\rm Vol}(\Omega_2) \wedge dx^6
\eea 
Like in \eqref{fd5} the metric here is also given in the string
frame. $A_{[2]}$ is the RR 2-form which couples to the D-string lying along
$x^6$ and $H_{[3]}$
is the NSNS magnetic 3-form field strength which couples to NS5-brane lying
along $x^1,\ldots,x^5$. Harmonic functions $H_{1,2}$ remain the same as given
above. Now taking T-dualities along the NS5 brane directions we generate all
the other NS5-D$p$ solutions which can be written in the following compact
form,
\bea\label{ns5dp}
ds^2 &=& H_1^{\half}H_2 \left[-\frac{dt^2}{H_1H_2}
+ \frac{\sum_{i=2}^p (dx^i)^2}{H_1H_2} + \frac{(dx^{1})^2}{H_1} + 
\frac{\sum_{j=p+1}^6
(dx^j)^2}{H_2} + dr^2 + r^2 d\Omega_{2}^2\right]\nn
e^{2\phi} &=& H_1^{\frac{3-p}{2}}H_2\nn
A_{[p+1]} &=& \left(1-H_1^{-1}\right)dt \wedge dx^1\wedge \ldots 
\wedge dx^{p}, \qquad H_{[3]}\,\,
=\,\, -Q_2 {\rm Vol}(\Omega_2) \wedge dx^1
\eea
Here $p = 1,\ldots,6$ and since these solutions are obtained from 1/4 BPS
threshold F-D5 solution by the application of S- and T-dualities, these
NS5-D$p$ solutions are also 1/4 BPS and threshold.
Note that in writing \eqref{ns5dp} from \eqref{ns5d1} we have exchanged the
coordinates $x^6 \leftrightarrow x^1$ for convenience. Now the D$p$-brane is
along $x^1,\ldots,x^p$ and NS5-brane is along $x^2,\ldots,x^6$ and their
charges are $Q_1$ and $Q_2$ respectively. They overlap on a $(p-1)$ brane
along $x^2,\ldots,x^p$. The harmonic functions $H_{1,2}$ remain the same as
given before. Next, we take a near horizon limit by approximating $H_{1,2}
\approx Q_{1,2}/r$ and substitute in the solution \eqref{ns5dp}. By 
introducing a new coordinate $u^2 = r$, we can write the NS5-D$p$ solution in
the near horizon limit\footnote{To further support the claim that gravity
does decouple for these solutions in the near horizon limit, beyond which
we have already discussed in footnote 5, we have studied the scattering
of a scalar field minimally coupled to the background \eqref{ns5dp}.
The dynamics, in the near horizon limit, is found to be described by a 
Schr\" odinger-like equation given as $(\partial_u^2  - V(u))\varphi(u) =0$.
Here $\varphi$ is related to the scalar field and the scattering potential is given 
as $V(u) = (-4Q_1Q_2 \omega^2 + 3/4 + 4\ell(\ell+1))/u^2$, where $\omega$
is the energy and $\ell=0,\,1,\,2, \ldots$ corresponds to various partial
waves. We thus find that for $\omega^2 < (3/16 + \ell(\ell+1))/(Q_1Q_2)$, the
scalar field experiences an infinite potential barrier at $u \to 0$ and thus 
gravity gets decoupled. However, at higher energies when $\omega^2$ exceeds 
the above limit, there is no potential barrier and the gravity may not decouple. 
The situation is very much like 
the decoupling in D5 or NS5 brane cases \cite{Alishahiha:2000qf}.} as follows, 
\bea\label{ns5dpnh}
ds^2 &=& Q_1^{\half}Q_2u \left[-\frac{dt^2}{Q_1Q_2}
+ \frac{\sum_{i=2}^p (dx^i)^2}{Q_1Q_2} + \frac{(dx^{1})^2}{Q_1u^2} + 
\frac{\sum_{j=p+1}^6
(dx^j)^2}{Q_2u^2} + 4\frac{du^2}{u^2} + d\Omega_{2}^2\right]\nn
e^{2\phi} &=& \frac{Q_1^{\frac{3-p}{2}}Q_2}{u^{5-p}}\nn
A_{[p+1]} &=& -\frac{u^2}{Q_1} dt \wedge dx^1\wedge \ldots \wedge dx^{p}, 
\qquad H_{[3]}\,\,
=\,\, -Q_2 {\rm Vol}(\Omega_2) \wedge dx^1
\eea
It is clear from \eqref{ns5dpnh} that under the scaling $t \to \lambda^0 t$,
$x^1 \to \lambda x^1$, $x^{p+1,\ldots,6} \to \lambda x^{p+1,\ldots,6}$, $u \to
\lambda u$, the part of the metric in the square bracket remains invariant.
But the full metric is not indicating that there is hyperscaling violation. In
order to find the hyperscaling violation exponent $\theta$, we dimensionally
reduce the metric on $S^2$  and also on $x^2,\ldots,x^p$ and express the 
reduced metric in Einstein frame which is given as,
\be\label{reduced}
ds_{9-p,E}^2 = Q_1^{\frac{2}{7-p}}Q_2u^{\frac{2(9-p)}{7-p}} 
\left[-\frac{dt^2}{Q_1Q_2} + \frac{(dx^{1})^2}{Q_1u^2} + \frac{\sum_{j=p+1}^6
(dx^j)^2}{Q_2u^2} + 4\frac{du^2}{u^2}\right]
\ee  
We therefore find that under the above scaling the reduced metric 
\eqref{reduced} transforms as 
\bea\label{transform}
ds_{9-p,E} &\to& \lambda^{(9-p)/(7-p)} ds_{9-p,E}\nn
&\equiv& \lambda^{\theta/d} ds_{9-p,E}
\eea
where $d=7-p$ is the spatial dimension of the boundary theory and
therefore, from \eqref{transform} we find the hyperscaling violation exponent 
to have the value $\theta=9-p$. Under the above scaling the other fields
transform as,
\be\label{transform1}
\phi \to \phi - \frac{5-p}{2}\log \lambda, \quad A_{[p+1]} \to \lambda^2
A_{[2]}, \quad H_{[3]} \to \lambda H_{[3]}
\ee
Therefore, we have shown that the NS5-D$p$ intersecting brane solutions in the
near horizon limit do give rise to Lifshitz-like space-time with $z=0$,
$\theta=9-p$ and $d=7-p$. It can be easily checked that these values of 
$(z,\,\theta,\,d)$ indeed satisfy the following NEC \cite{Dong:2012se},
\bea\label{NEC}
(d-\theta)(d(z-1)-\theta)\geq 0\nn
(z-1)(d+z-\theta)\geq 0
\eea
Similar Lifshitz-like structures have also been obtained for near horizon
geometries of F-D$p$ \cite{Dey:2012tg} and also in some intersecting brane 
solutions \cite{Dey:2012rs}. We here
give a table for the values of $(z,\,\theta,\,d)$ of all these solutions for
comparison,

\vspace{.3cm}

\begin{center}
\begin{tabular}{||c|c|c|c||}
\hline\hline
Type & $z$ & $\theta$ & $d$ \\ \hline\hline
\hfil & \hfil & \hfil & \hfil \\
NS5-D$p$ & 0 & $9-p$ & $7-p$ \\ 
$1\leq p \leq 6$ & \hfil & \hfil & \hfil \\ \hline
\hfil & \hfil & \hfil & \hfil \\
F-D$p$ & $\frac{2(5-p)}{4-p}$ & $p - \frac{p-2}{4-p}$ & $p+1$ \\
$0\leq p \leq 5,\, p\neq 4$ & \hfil & \hfil & \hfil \\ \hline
D0-D4 & \hfil & \hfil & \hfil \\
D1-D3 & 4 & 2 & 4 \\
D2-D2$^\prime$ & \hfil & \hfil & \hfil \\ \hline
D2-D6 & \hfil & \hfil & \hfil \\
D3-D5 & 0 & 6 & 4 \\
D4-D4$^\prime$ & \hfil & \hfil & \hfil \\ \hline\hline
\end{tabular}
\end{center}
\vspace{.3cm}

In the third row, the three intersecting brane solutions D0-D4, D1-D3 and 
D2-D2$^\prime$ have the same set of values of $(z,\,\theta,\,d)$ and
similarly, in the 4th row D2-D6, D3-D5 and D4-D4$^\prime$ have the same set of
values of $(z,\,\theta,\,d)$. Among all the solutions only for F-D2 we have
$\theta = d-1$ indicating that the corresponding boundary theory describes
compressible metallic states with hidden Fermi surface \cite{Ogawa:2011bz,
Huijse:2011ef}. The field theoretic meaning of the other scaling solutions 
are not known.

\section{RG flows and phase structures}
  
We have seen in the previous section that the dilatons of NS5-D$p$ solutions
are in general not constant except for $p=5$. Therefore, as $u$ changes, the
dilatons can become large and invalidate the supergravity solutions. Also the
curvature of the solutions must remain small for the
supergravity description to remain valid. These conditions put some
restrictions on the parameter $u$ or the energy parameter in the dual field
theory. We will discuss them here in a case by case basis.  

\subsection{NS5-D1}
 
The S-dual of this solution is F-D5 and this case is already discussed
in \cite{Dey:2012tg} and will not be repeated here. Note that both these
solutions have Lifshitz-like structure in the near horizon limit.

\subsection{NS5-D2}

The near horizon geometry in a suitable coordinate $u$ of this solution is 
given
in \eqref{ns5dpnh}. From there we find that the dilaton remains small only if
$u \gg Q_1^{1/6}Q_2^{1/3}$ and the curvature remains small if $u \gg
1/(Q_1^{1/2}Q_2)$. For large $Q_1$ and $Q_2$ if the first condition is
satisfied then the second one is automatically satisfied. However, when $u
\leq Q_1^{1/6}Q_2^{1/3}$, the dilaton would be large and in order to get a
valid supergravity description we need to uplift the solution to M-theory.
The solution in this case takes the form,
\be\label{ns5d2m}
ds^2 = Q_1^{\frac{1}{3}} Q_2^{\frac{2}{3}}\left[\frac{-dt^2 + (dx^2)^2}{Q_1Q_2
    r^2} + \frac{(dx^1)^2}{Q_1} + \frac{\sum_{j=3}^6(dx^j)^2}{Q_2} +
  \frac{dr^2}{r^2} + d\Omega_3^2\right]
\ee
The above solution represents the near horizon limit of M5-M2 solutions
intersecting on a string along $x^2$ \cite{Tseytlin:1997cs}. Note that in 
writing the solution we
have kept the coordinate $r$ (not $u$). It is obvious that it has an AdS$_3$
$\times$ E$^5$ $\times$ S$^3$ structure. 

\subsection{NS5-D3}

The S-dual of this solution is F-D3 and this case is also discussed in
\cite{Dey:2012tg} and will not be repeated here. Here also both the solutions
have Lifshitz-like structure.

\subsection{NS5-D4}
 
The geometry of this solution in the near horizon limit is given in
\eqref{ns5dpnh}. From the expression of dilaton we find that it will remain
small for $u\gg Q_2/Q_1^{1/2}$ and the curvature remains small for 
$u \gg 1/(Q_1^{1/2}Q_2)$. For large $Q_2$, the first condition is sufficient for
supergravity description to remain valid. However, when $u \leq
Q_2/Q_1^{1/2}$, we have to uplift the solution to M-theory. The uplifted
solution has the form,
\be\label{ns5d4m}
ds^2 = (Q_1Q_2)^{\frac{2}{3}} u^{\frac{4}{3}}\left[\frac{-dt^2 + \sum_{i=2}^4
(dx^i)^2}{Q_1Q_2} + \frac{(dx^1)^2}{Q_1 u^2} + 
\frac{\sum_{j=5}^6(dx^j)^2}{Q_2u^2} + 4\frac{du^2}{u^2} + d\Omega_2^2
+ \frac{(dx^{11})^2}{Q_1u^2}\right]
\ee  
The above solution represents the near horizon limit of M5-M5$^\prime$
solution intersecting on a 3-brane along $x^2,\,x^3,\,x^4$ 
\cite{Tseytlin:1997cs}. This metric again
has a Lifshitz-like structure with $(z=0,\,\theta=6,\,d=4)$. In obtaining
$\theta$ we compactify \eqref{ns5d4m} on S$^2$ and $x^{2,3,4}$ and
express the resulting metric in Einstein frame.

\subsection{NS5-D5}

In the near horizon limit NS5-D5 solution expressed in a suitable coordinate
$u$ is given in \eqref{ns5dpnh}. Here the dilaton is constant. However, the
dilaton will remain small as long as $Q_2 \ll Q_1$ and the curvature remains
small for $u \gg 1/(Q_1^{1/2}Q_2)$. But, when $Q_2 \geq Q_1$, the dilaton will
become large invalidating the supergravity description. In that case we have
to go to the S-dual frame. In the S-dual frame the metric takes the form,
\be\label{ns5d5s}
ds^2 = Q_1Q_2^{\frac{1}{2}} u \left[\frac{-dt^2 + \sum_{i=2}^5
(dx^i)^2}{Q_1Q_2} + \frac{(dx^1)^2}{Q_1 u^2} + 
\frac{(dx^6)^2}{Q_2u^2} + 4\frac{du^2}{u^2} + d\Omega_2^2\right]
\ee 
Again we find that the above S-dual metric has Lifshitz-like structure with
$(z=0,\,\theta=4,\,d=2)$ exactly as those of the original solution.

\subsection{NS5-D6}

The near horizon limit of this solution in a suitable coordinate $u$ is given
in \eqref{ns5dpnh}. Here the dilaton and the curvature of the metric remain
small in the region $1/(Q_1^{1/2} Q_2) \ll u \ll Q_1^{3/2}/Q_2$. However, when
$u \geq Q_1^{3/2}/Q_1$, the dilaton becomes large and the supergravity
solution is no longer valid. So, we have to uplift the solution to
M-theory. The uplifted solution in this case takes the form,
\be\label{ns5d6m}
ds^2 = Q_1Q_2^{\frac{2}{3}} u^{\frac{2}{3}} \left[\frac{-dt^2 + \sum_{i=2}^6
(dx^i)^2}{Q_1Q_2} + \frac{(dx^1)^2}{Q_1 u^2}
+ 4\frac{du^2}{u^2} + d\Omega_2^2 + \frac{(dx^{11} - 2 Q_1\sin^2 
\frac{\theta}{2}d\phi)^2}{Q_1^2} \right]
\ee  
The above solution represents the near horizon limit of the intersecting
M5-KK solution \cite{Bergshoeff:1997tt}. It can be easily checked that 
this solution also has
Lifshitz-like structure with $(z=0,\,\theta=3,\,d=1)$ as those of the original
solution. 

\section{Conclusion}  

To conclude, in this paper we have constructed 1/4 BPS, threshold intersecting
NS5-D$p$ (for $1\leq p \leq 6$) brane solutions of type II string theories 
starting from the 1/4 BPS
threshold F-D5 solution given in \cite{Dey:2012tg} of type IIB string theory 
by applying S- and T-dualities.  
These solutions are different from the known NS5-D$p$ solutions which are 1/2
BPS and non-threshold. 
Unlike the known NS5-D$p$ solutions which in the near
horizon limit leads to six dimensional open D$p$-brane theory on the boundary,
the solutions constructed in this paper give Lifshitz-like space-time, that
is, space-time having Lifshitz scaling symmetry along with some hyperscaling
violation in the near horizon limit. It is known that some condensed matter 
system near their quantum critical point show similar kind of scaling 
symmetry, so, the solutions discussed in this paper may describe the 
gravity dual of such condensed matter
systems near that point. Also since the NS5-D$p$ solution we have obtained are
1/4 BPS, the Lifshitz-like solutions should also preserve at least a 1/4 
SUSY. We found that the solutions obtained from NS5-D$p$ have a scaling
symmetry with dynamical critical exponent $z=0$, the hyperscaling violation
exponent $\theta=9-p$ and the spatial dimensions of the boundary theory
$d=7-p$. We have seen that these values of $(z,\,\theta,\,d)$ satisfy NEC
indicating that they might lead to a sensible dual theory. As similar
structures were obtained for other type of intersecting solutions we have
given a table for the comparison. The dilatons for these solutions are
not constant except for $p=5$ and therefore, they give rise to RG flows. 
We have discussed
the various phases of these solutions considering the RG flows. We found that
the strongly coupled phases also give Lifshitz-like
space-time except for $p=2$. The strongly coupled phase of the near horizon 
NS5-D2 solution actually has the structure AdS$_3$ $\times$ E$^5$ 
$\times$ S$^3$. The scaling solutions we have obtained in this paper 
has dynamical scaling exponent $z=0$ and thus it appears that there is 
no relaxation in time of the system described by the boundary theory. 
It would be interesting to understand the field theoretic meaning and 
the consequences of these scaling solutions.       
     
\section*{Acknowledgements}

One of the authors (PD) would like to acknowledge thankfully the financial
support of the Council of Scientific and Industrial Research, India
(SPM-07/489 (0089)/2010-EMR-I).

\end{document}